# Development and validation of a short form of the medication literacy scale for Chinese College Students

Zhenzhen Chen Bachelor Candidate[1] |& Jiabao Ren Master Candidate[2] |(the authors contribute equally) |Tingyu Duan Master Candidate[3] |Ke Chen Master Candidate[4] |Ruyi Hou Master Candidate[6] |Yimiao Li Master Candidate[6]|Leixiao Zeng PhD student[5] |Xiaoxuan Meng Master Candidate[6] |Yibo Wu, PhD candidate, MPharm[7] |Yu Liu, PhD,Prof[2].

**Background:** Medication literacy is an essential component of health literacy, and its level is a crucial indicator of medication safety and adherence. Medication literacy is defined as an individual's ability to recognize, understand, and communicate medication information. However, these scales generally require a longer assessment time and are mostly applicable to patients and community residents, with a limited scope of application. Therefore, it is necessary to development of a short form of the medication literacy scale.

[1] College of Science, Minzu University of China, Beijing, China

[2] School of Nursing, China Medical University, Shenyang, Liaoning Province, China

[3] Hebei Institute Of Communications,Hebei, China

[4] Department of Social Science and Humanities, Harbin Medical University, Harbin, Heilongjiang Province, China

[5] School of Journalism and Communication,Renmin University of China,Beijing ,China

[6] Tianjin Medical University,Tianjin, China

[7] School of Public Health, Peking University, Beijing,China

Correspondence 1: Yibo Wu, School of Public Health, Peking University,Beijing, China.Email: bjmuwuyibo@outlook.com

Correspondence 2: Yu Liu, School of Nursing, China Medical University,Shenyang, Liaoning Province, China.

**Objective:** To develop a short version of the Medication Literacy Scale (MLS-14) based on classical test theory (CTT) and item response theory (IRT) and conduct psychometric testing in a college student population.

**Methods:** The current study developed a short version of the MLS-14 and recruited 2431 Chinese college students as participants to analyze its psychometric characteristics. a 6-item version of the scale (MLS-SF) was simplified based on classical test theory(CTT)and item response theory(IRT) methods.

**Results:** Using Classical Test Theory and Item Response Theory, a 6-item scale (MLS-SF) version was simplified, with a Cronbach's α coefficient of 0.765. The MLS-SF extracted three common factors through exploratory factor analysis, with a cumulative variance contribution rate of 66%, and all item factor loadings were >0.5. The simplified scale (MLS-SF) showed good fit indices about the three-factor structure in confirmatory factor analysis ($\chi^2/df=5.11$, RMSEA=0.063, GFI=0.990, AGFI=0.966, NFI=0.984, IFI=0.987, CFI=0.987). According to the IRT model, all 6 items had reasonable discrimination parameters and location parameters and were DIF-free by gender. Except for Items 4 and 10, all the other items had high information at medium $\theta$ levels.

**Conclusion:** The 6-item Medication Literacy Short Form (MLS-SF) developed in this study has good reliability and validity and is an effective tool for quickly assessing the medication literacy of college students.

# Introduction

Medication literacy (ML) is an important component of health literacy and an important factor in ensuring medication safety and adherence[1]. In March 2017, the World Health Organization (WHO) launched the third Global Patient Safety Challenge with the theme of medication without harm, which aims to gain worldwide commitment and action to reduce severe, avoidable medication-related harm by 50% over the five years to 2024[2]. Inadequate medication literacy can increase the risk of medication errors, which will harm persons' health, by leading to poor adherence and misunderstanding of medication-related information or instructions[3]. Medication literacy directly affects an individual's medication behavior, medication safety, and medication effects, which in turn affects the individual's health status and quality of life[4-6].

Medication Literacy is defined as the ability of individuals to obtain, comprehend, communicate, calculate, and process patient-specific information about their medications. This enables them to make informed decisions about their health and medication use, regardless of the mode of delivery (e.g. written, oral, or visual) [7]. Medication literacy is a type of health literacy that focuses on the ability to review, communicate, and process information about medications[8]. People with lower medication-related health literacy may have difficulties in understanding medication labels, instructions, warnings, and advertisements[9, 10]. Previous studies have shown that college students have low levels of medication literacy and often engage in unsafe and irrational medication behaviors, such as self-medication, non-adherence, misuse, abuse, and overdose[4, 11, 12]. These behaviors may lead to adverse drug reactions, drug interactions, drug resistance, and other health problems[13].

The purpose of conducting Item Response Theory (IRT) analysis is to examine the functional characteristics of items, such as item difficulty and item discrimination [14] At present, a variety of medication literacy scales are available around the world for use in different populations and scenarios, such as hypertensive patients, coronary heart disease patients, and college students[1, 15-17]. Several tools for measuring medication literacy (ML) have been developed and used in research studies. However, these scales have some common shortcomings, such as too many items, repetitive content, complex structure, cumbersome operation, long time consumption, etc.which affect the efficiency and effectiveness of the application of the scales[17]. Therefore, it is necessary to simplify the existing medication literacy scales to improve their operability and reliability on CTT (Classic test theory)and IRT (item response theory). The use of CTT is grounded in its long-standing history and widespread recognition within the psychometric community, providing a robust framework for assessing the reliability, validity, and other key psychometric properties of a test. It offers a comprehensive system for calculating these parameters, thereby enhancing the objectivity of testing[18]. However, CTT has its limitations, including sample dependency and a lack of precision in measuring individual variability. This is where IRT offers a complementary advantage. IRT provides sample-invariant parameter estimates, allowing for more precise measurements tailored to individual levels and the ability to accurately estimate item and test information independent of the sample[19]. Integrating both CTT and IRT in tandem enhances the psychometric evaluation of measurement tools, allowing for a comprehensive assessment of reliability, validity, and individualized measurement precision[20, 21].

In light of this, the present research aimed to develop a short version of Medication Literacy Scale (MLS-SF) which would be tested for reliability and validity in a wider range and further simplified in this study, a short form yielding satisfactory results that simple, easy to complete, and not time-consuming.

## Methods

**2.1 Participants**

In this study, 1,323 college students from Changzhi Medical College[22] and 1,108 college students from more than a dozen universities in Shanxi Province [23](see Table I) were recruited to complete the questionnaire. After excluding the invalid questionnaires from 84 college students from Changzhi and 75 college students from Shanxi Province due to short (<60 s) and long (>2400 s) answer times and the presence of obvious logical fallacies. In Sample I, 1239 valid questionnaires from Changzhi Medical College were included (mean age: 21.06, SD = 1.04, CI = 95%) with a valid response rate of 93.65%. In sample II, 1033 valid questionnaires from other universities in Shanxi province were included (mean age 20.38, SD=1.148, CI=95%) with a valid response rate of 93.23%.

## 2.2 Procedure

Sample I and Sample II used the Questionnaire Star platform to develop an online questionnaire that included a medication literacy scale, medication habits, access to medication knowledge, and some demographic characteristics. The questionnaire was distributed from March to June 2020 to college students without cognitive impairment from more than ten universities in Shanxi Province. Each participant was informed of the purpose of the questionnaire, the confidentiality format, and the estimated time spent prior to participation.

## 2.3 Measures

### 2.3.1 Demographic information

Demographic information included age, sex, major, cost of living, marital status, parental education, and location.

### 2.3.2 The Medication Literacy Scale (MLS-14)

The original Medication Literacy Scale (MLS-14)[23] served as the basis for this study, which adopted the 14-item health literacy scale (HLS-14) for Japanese adults[24], based on the structural equation model (SEM),was utilized to assess ML levels. The scale comprises three dimensions: functional ML (5 items), communicative ML (5 items), and critical ML (4 items). Responses were based on a 5-point Likert scale ranging from "strongly disagree" to "strongly agree", from which the five items in the functional ML dimension are reverse scored. The total ML score ranged from 14 to 70, with higher scores indicating better ML. Moreover, the scale has shown high internal consistencies (Cronbach's alpha values 0.831). In this study we denote the items 1-5 of functional ML dimension as A1-A5, items 6-10 of communicative ML as B1-B5, and items 11-14 of critical ML as C1-C4.

## 2.4. Data analysis

### 2.4.1 CTT analysis

The structure of the Medication Literacy Scale (MSL-14 ) was measured using an exploring factors analysis,(EFA) which was performed using the principal component analysis with varimax rotation, the number of factors was determined according to the theoretical structure at the time of construction of the scale, eigenvalues greater than 1 were extracted from the common factors, and items with factor loadings <0.4 were deleted.

Calculate the Pearson correlation coefficient between each item and the total score of the scale, and select those items whose absolute value is large and statistically significant[25].

The Cronbach's α coefficient of the total scale or each item was calculated. If the Cronbach's alpha coefficient of the total scale or each item increased after deleting an item, it indicated that the existence of the item reduced the internal consistency of the scale or dimension and should be deleted. Otherwise, the item should be retained[26].

The total score of the scale was ranked in descending order, and the 27% of the total score at the high and low ends of the scale were taken as the boundaries, and an independent samples t-test was used to compare the differences in the scores of the items in the first 27% of the total score (the high subgroup) with those in the last 27% (the low subgroup), and the item with no statistically significant difference indicated that it had poor discriminatory power, and then that item should be deleted[27].Floor or ceiling effects were used to exclude items with limited content validity. An item should be removed if floor or ceiling effects exceed 15%. According to Floor or ceiling effects, A1,A2,B1 were deleted[28].

**2.4.2 IRT analysis**

Multilevel score models for IRT are the generalized partial credit model (GPCM)[29] and the graded response model (GRM). The lower the Akaike Information Criterion (AIC) and Bayesian Information Criterion (BIC) values, the better the model fit [29]. Based on the fit indices of AIC, BIC, and -2Log-Likelihood, this study compared the fit of the Measurement of Medication Literacy Scale (MLS-14) under the Grading Response Model (GRM) and the Generalized Partial Credit Model (GPCM). Model fit was evaluated by the sum-score-based item fit statistic (S-$X^2$). items flagged by these tests below a p-value level of 0.05 were considered as potentially problematic items [30].

In the GRM, items are described in terms of a slope parameter (also called discrimination parameter and often denoted by α and category thresholds (denoted by b). Items with higher slopes offer better discrimination between those with high and low score levels on the ML dimension assessed by the items. In the GRM, category thresholds indicate for each category, the locations on the latent scale below which respondents would tend to choose that particular category or worse, rather than the categories indicating better literacy of medication Hence, they are indicative of the graduated nature ('severity') of the items and provide useful information on the coverage in terms of contribution to measurement precision at different locations across the latent scale[31]. Each item was estimated to ensure that the retained items have a discrimination value greater than 0.35, and the range of the four difficulty values ($b_1$, $b_2$, $b_3$, and $b_4$) should be between -3 and 3[32].

Information Characteristic Curve (ICC) and Item Information Curve (IIC) are important tools for assessing item quality. ICC values ranged from 0 to 1. Better repeatability was indicated by a higher value. ICC values are grouped into four classes: inferior (less than 0.40), moderate (between 0.41 and 0.60), good (between 0.61 and 0.80), and excellent (more than 0.80)[33]. In addition, by plotting the item information curves, by comparing the IICs of different items, it is possible to select the items that contribute the most to the estimation of the tester's ability. In general, questions with higher peaks and wider information bandwidths are considered more useful[34]. By combining these two, items with high information value and/or good information distribution within the target ability range were retained, while items with low information value, sub-optimal ICC, or high overlap with other items were removed.

In this study, Logistic regression (LR) was used to calculate CHISQR for DIF testing of age, gender, and ethnicity. If the chi-square value is large and the corresponding p-value is less than the pre-set significance level (such as 0.05), then it can be considered the presence of DIF threatens the fairness and validity of the scale. Therefore, removing the items that exhibited DIF.

### 2.4.3. Content analysis

The results of statistical analyses of items that meet the initial screening requirements are evaluated in conjunction with expert opinions to determine the final retained items. Seven experts were invited to evaluate the content validity of the scale using a Likert 4-point scale, with 1 indicating "not relevant", 2 indicating "weakly relevant", 3 indicating "strongly relevant", and 4 indicating "very relevant". A score of 1 means "not relevant", 2 means "weakly relevant", 3 means "strongly relevant", and 4 means "very relevant". Based on the results of the expert scores, the content validity index (S-CVI/UA) and the content validity index (I-CVI) of the items were calculated, and if the values of S-CVI/UA and ICVI were both ≥0.800, it indicated that the content validity of the scale was good.

2.5. **Data Analysis Software and Procedure**

All data manipulations were conducted using R version 4.3.3, IBM SPSS Statistics 26. (IBM Corporation, Armonk, NY, USA) and Amos Graphic 28.0 (IBM, New York, NY, USA. Before proceeding with the formal data analysis, we utilized IBM SPSS Statistics 26 to obtain the socio-demographic information of the two datasets. This included counts and percentages for reported categorical variables, as well as means and standard deviations for continuous variables, to better understand the basic characteristics of each sample (as illustrated in the figures). Initially, we tested the psychometric indicators of the MLS-14 using sample 1. Subsequently, based on exploratory analysis, we conducted confirmatory analysis on the derived abbreviated scale using sample 2, to cross-validate our findings [27, 35]. Thereafter, the reliability and structural validity of the simplified scale were also measured. Psychometric testing of the full and short version of the scale used IBM SPSS Amos 28.0 to analyze.

The remaining analyses were performed using packages in R: the *epiDisplay* package was used for Cronbach's alpha, the *psych* package was used for factor analysis, the *mirt* package was used for the fit analysis of the IRT model, for estimating IRT parameters and for plotting item characteristic curves ICCs. The *catR* package was used to draw item information curves and test the information function. The *lordif* package was used to identify DIFs.

# Result

### 3.1. Socio-demographic information on the study population

In Sample 1, 379 (30.59%) are males and 860 (69.41%) are females; 40.68% are from urban areas and 59.32% are from rural areas; 0 (0%) were freshmen, 481 (38.82%) were sophomores, 346 (27.93%) were juniors, 266 (21.47%) were seniors and 145 (11.70%) were seniors; 1 (1.0%) was a graduate or above; and 1 (1.0%) was a graduate or above. 1 person above. ; 26 (2.04%) in Information Management and Information Systems, 23 (1.74%) in Communication, 56 (6.65%) in Science, 16 (1.21%) in Art, 74 (5.75%) in Pharmacy, and 1044 (82.62%) in Medicine. In sample two. 368 males (35.62 %) and 647 females (62.63 %); 47.92 % were from towns and 52.080 % were from rural areas; 294 majors in literature, 184 majors in engineering, 143 majors in economics and management, 115 majors in science, 54 majors in pedagogy, 55 majors in law, 79 majors in agronomy, 23 majors in military science and 86 majors in medicine.

**Table1** Comparisons of basic characteristics between the training group and validation group

|  | Development sample(N=1239) | Validation sample(N=1033) | $\chi^2/t$ | $P$ |
|---|---|---|---|---|
| Gender(n,%) |  |  | 8.086[b] | 0.003 |
| Male | 379(30.59%) | 368(35.62%) |  |  |
| Female | 860(69.41%) | 647(62.63%) |  |  |
| Education(n,%) |  |  | 553.171[b] | <0.001 |
| First-year undergraduate | 0(0%) | 220(21.30%) | | |
| Second-year undergraduate | 481(38.82%) | 395(38.24%) | | |
| Third-year undergraduate | 346(27.93%) | 281(27.20%) | | |
| Fourth-year undergraduate | 266(21.47%) | 100(10.33%) | | |
| Fifth-year undergraduate | 145(11.70%) | 6(0.85%) | | |
| Postgraduate | 1(0.80%) | 122(11.81%) | | |
| Marital status(n,%) | | | 16.211[b] | 0.001 |
| Single or unmarried | 1232(99.44%) | 1010(97.77%) | | |
| Separation or divorce | 4(0.32%) | 3(0.29%) | | |
| Married | 2(0.16%) | 11(1.07%) | | |
| Widowed spouse | 1(0.08%) | 9(0.87%) | | |
| Place of residence (n,%) | | | 11.988[b] | 0.001 |
| Rural | 50440.68(%) | 495(47.92%) | | |
| Urban | 735(59.32%) | 538(52.08%) | | |
| Nation(n,%) | | | 31.099[b] | <0.001 |
| Han | 120797.42(%) | 954(92.35%) | | |
| Minority | 32(2.58%) | 79(7.65%) | | |
| Total Score($\bar{X} \pm SD$) | 48.05±7.02 | 46.01±6.32 | 7.22[a] | <0.01 |

[a] *t*-test, [b] Chi-squared test

Abbreviations: SD, Standard Deviation.

### 3.2. Psychometric testing of the full version of the scale

Based on the data from Sample1, a confirmatory factor analysis of the psychometric indicators of the full version of the 14-item MLS showed a good model fit ($\chi^2$/df =5.759, NFI=0.928, CFI=0.939, RMSEA=0.069). The fit indicators showed that the original scale model was well fitted. Calculation of the internal consistency coefficient yielded Cronbach's$\alpha$= 0.831, with good scale reliability.

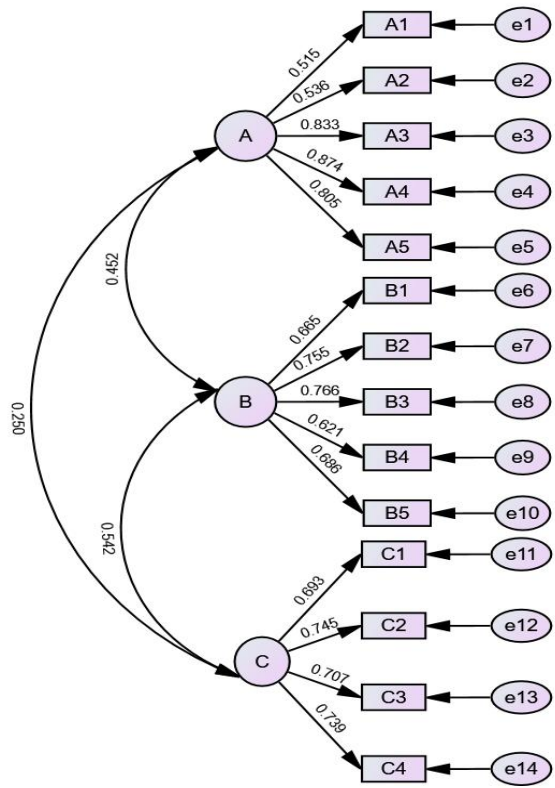

Figure 1 Figure 3 The Medication Literary Scale : Confirmatory factor analysis : Three-factor model :A: functional ML, B:communicative ML, and C:critical ML

### 3.3. Classic Test Theory (CTT)

To ensure the randomness of the data, a random sample of 1000 data was set up. Using cutoffs suggested by Hu and Bentler[36],the Bartlett's Spherical Test and KMO measure were performed on the data based on Sample 1, and the Bartlett's Spherical Test ($p$<0.01) and the KMO measure was good (0.88), suitable for factor analysis. Exploratory factor analysis was conducted on the Medication Literacy Scale using principal component analysis and varimax rotation. Kaiser's [37]criterion, involving retaining factors with eigenvalues greater than one, suggested a unidimensional factor solution, and the parallel analysis scree plot(Figure 2) supported this factor solution, which extracted three factors with eigenvalues greater than 1[33],which accounted for 61% of the total variance. Factor loadings were then measured for each item in each dimension, and the items with the two highest factor loadings for each dimension were retained.

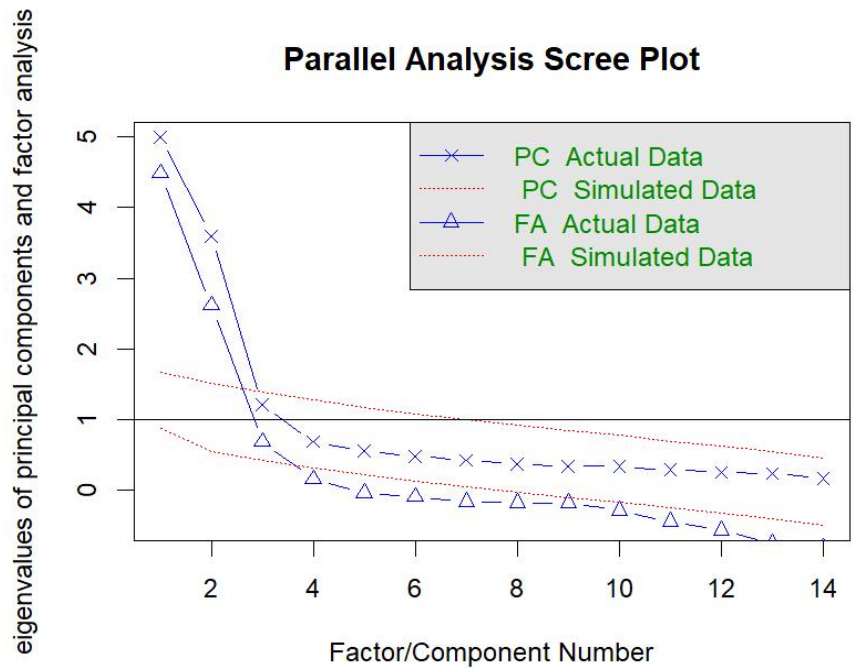

Figure2 Parallel Analysis Scree Plot

The correlation coefficient method was used for item analysis. The results showed that the correlation coefficients between the original version of the medication Literacy Scale and each of the items were $r > 0.77$ (value range:0.778-0.894), indicating good agreement between the items and the dimensions, and the two items with the highest correlation coefficients for each dimension were retained.

The Cronbach's $\alpha=0.831$ for the original version of the MLS-14 was obtained based on Sample 1, and the Cronbach's $\alpha$ decreased(value range:0.814~0.829)after removing each item. The least two items in each dimension that have the smallest decreased in internal consistency after removal were considered to be retained.

The highest 27% (⩾51 points) and the lowest 27% (⩽44 points) of the total score of the Medication literacy scale were divided into high and low subgroups, and further independent samples t-tests were conducted, which showed that there was a significant difference between the scores of high and low subgroups of the scale on each item ($P<0.001$).

According to Floor or ceiling effects A1, A2, B1were deleted. The specific results are shown in Table 1.

Finally, by combining the retained scales of the above four methods, the six items with the highest number of retained scales. The simplified medication Literacy Scale short form (MLS-SF) based on classical test theory consists of a total of 6 items: A3, A4, B2, B3, C2, and C4, respectively.

**Table 2** Classic Test Theory (CTT): MLS-14 Item descriptive and correlation results (reversed scored).

| Item | Floor effect (%) | Ceiling effect (%) | Correlation Coefficient | CTIC | Independent sample t-test | Factor Loading | Number of selected | Reserved Items |
|---|---|---|---|---|---|---|---|---|
| A1 | 1.5 | 21.1 | 0.778 | 0.892 | <0.001 | 0.67 | 3 | |
| A2 | 1.5 | 20.9 | 0.780 | 0.827 | <0.001 | 0.68 | 3 | |
| A3 | 2.7 | 13.8 | 0.889 | 0.820 | <0.001 | 0.88 | 5 | √ |
| A4 | 2.5 | 12.1 | 0.894 | 0.820 | <0.001 | 0.90 | 5 | √ |
| A5 | 3.5 | 10.3 | 0.867 | 0.822 | <0.001 | 0.85 | 4 | |
| B1 | 15 | 2.1 | 0.796 | 0.820 | <0.001 | 0.69 | 3 | |
| B2 | 11.2 | 1.3 | 0.838 | 0.816 | <0.001 | 0.78 | 5 | √ |
| B3 | 10.2 | 1.3 | 0.842 | 0.814 | <0.001 | 0.79 | 5 | √ |
| B4 | 8.7 | 1.9 | 0.791 | 0.817 | <0.001 | 0.65 | 4 | |
| B5 | 8.6 | 2.7 | 0.817 | 0.815 | <0.001 | 0.71 | 3 | |
| C1 | 13.2 | 1.3 | 0.823 | 0.821 | <0.001 | 0.67 | 3 | |
| C2 | 5.1 | 1.1 | 0.861 | 0.823 | <0.001 | 0.79 | 6 | √ |
| C3 | 9.3 | 1.8 | 0.848 | 0.822 | <0.001 | 0.70 | 4 | |
| C4 | 7.2 | 2.4 | 0.848 | 0.820 | <0.001 | 0.71 | 5 | √ |

Note: Factor analysis: the loading of the item on the corresponding common factor;

correlation coefficient: the correlation coefficient between the item and the score of the aspect to which it belongs; "√" indicates that the item was selected for the final scale. Bold word indicated values did not meet standard.
Abbreviations: CITC, Corrected Item-total Correlation

**3.3** Item Response Theory (IRT)

Based on the IRT analysis results, we selected items with high discrimination and moderate difficulty, while excluding items with differential item functioning (DIF) based on gender, place of origin, ethnicity, and items that lead to good model fit. Items with high information value and/or good information distribution within the target ability range were retained, ultimately constructing a simplified medication literacy scale (MLS-SF) as shown in the figure below.

Considering the lower AIC, BIC, and -2Log-Lik values of the GRM model, which indicates a better fit. (GRM: AIC=30269, BIC=30613, -2Log-Lik=30128. GPCM: AIC=30519, BIC=30864, -2Log-Lik=30380). Therefore, in this study, we adopted the Graded Response Model (GRM) in IRT, which assumes that participants' responses to each item are ordered and each item's response category is related to the participants' level of medication literacy. We calculate the $S-X^2$ statistic to assess the model fit of each item and deletes items with poor fit ($P<0.05$)[38], which includes items A1, A2, and B1. The items of MLS-14 all have high discrimination parameters, ranging from 1.27 to 4.17, with reasonable and extensive location parameters, indicating that these items can distinguish individuals with different levels of pharmaceutical literacy. However, the difficulty parameters (b) of items A1 and C1 are not within the range of -3 to +3, so they are considered for deletion.

**Table 3** Multidimensional parameters of discrimination and difficulty from the full information confirmatory graded response model (GRM) (MLS-14)

| Item | Dimension | Category Thresholds | | | | | Result |
|---|---|---|---|---|---|---|---|
| | | Slope α | $b_1$ | $b_2$ | $b_3$ | $b_4$ | |
| A1 | Functional ML | 1.27 | -3.03 | -0.285 | 1.0876 | 3.07 | × |
| A2 | | 1.28 | -1.93 | 0.206 | 1.4551 | 2.89 | |
| A3 | | 3.40 | -2.15 | -0.886 | 0.3504 | 1.89 | |
| A4 | | 4.17 | -1.99 | -0.754 | 0.3416 | 1.86 | |
| A5 | | 3.02 | -2.13 | -1.765 | 0.3224 | 1.74 | |
| B1 | Communicative ML | 2.18 | -2.60 | -1.507 | 0.4185 | 1.72 | |
| B2 | | 3.04 | -2.61 | -1.478 | -0.3533 | 1.69 | |
| B3 | | 3.01 | -2.60 | -1.539 | -0.1464 | 1.79 | |
| B4 | | 1.82 | -2.75 | -1.236 | -0.991 | 2.12 | |
| B5 | | 2.23 | -2.62 | -1.346 | -0.0047 | 2.04 | |
| C1 | Critical ML | 2.85 | -3.03 | -1.926 | -0.9575 | 1.34 | × |
| C2 | | 2.99 | -2.72 | -1.577 | -0.6679 | 1.44 | |
| C3 | | 2.48 | -2.48 | -1.424 | -0.3792 | 1.67 | |
| C4 | | 2.58 | -2.79 | -1.622 | -0.5786 | 1.51 | |

Note: '√' represented the selected item; '×' indicated the item considered to be deleted; α, discrimination parameter; $b_{1-3}$, difficulty parameters; Bold word indicated values did not meet standard.

Areviations: MLS-14, Medication Literary Scale

The information values of items A1 and A2 on the first dimension show a smaller performance(<0.5) at the θ level, and the information of items B4 below 1 at the θ level and B5 below 1 between 0-2 on the second dimension is also small; therefore, items A1, A2, and B4, B5 are deleted. (online supplement)

Items A1, A2, B1, B4, C1, and C3 show different degrees of project functional differences in terms of gender, place of origin and ethnicity, and are considered for deletion. For the categorical variable of gender, items A2 and C3 show item functioning differences between males and females. For the categorical variable of place of origin, items B1 and B4 show item functioning differences between urban and rural students. For the categorical variable of ethnicity, items A1 and C3 show item functioning differences between the Han majority and other ethnic groups. All other items meet the above requirements and are therefore retained. ( Online Supplement )

Finally, based on the IRT analysis results, we selected items with high discrimination and moderate difficulty, while excluding items with differential item functioning (DIF) based on gender, place of origin, ethnicity, and items that lead to good model fit. Items with high information value and/or good information distribution within the target ability range were retained. Thus, through the IRT method, 7 items are retained, namely A3, A4, A5, B2, B3, C2, and C4.

### 3.5 Construction of medication literary short form scale

Based on the CTT and IRT analysis results, this study adopts the "majority voting method" to maintain the structure and information quantity of the original scale. Two items with better quality are selected for the functional medication literacy (ML) dimension, two for the communicative ML dimension, and two for the critical ML dimension, ultimately forming a 6-item simplified scale.

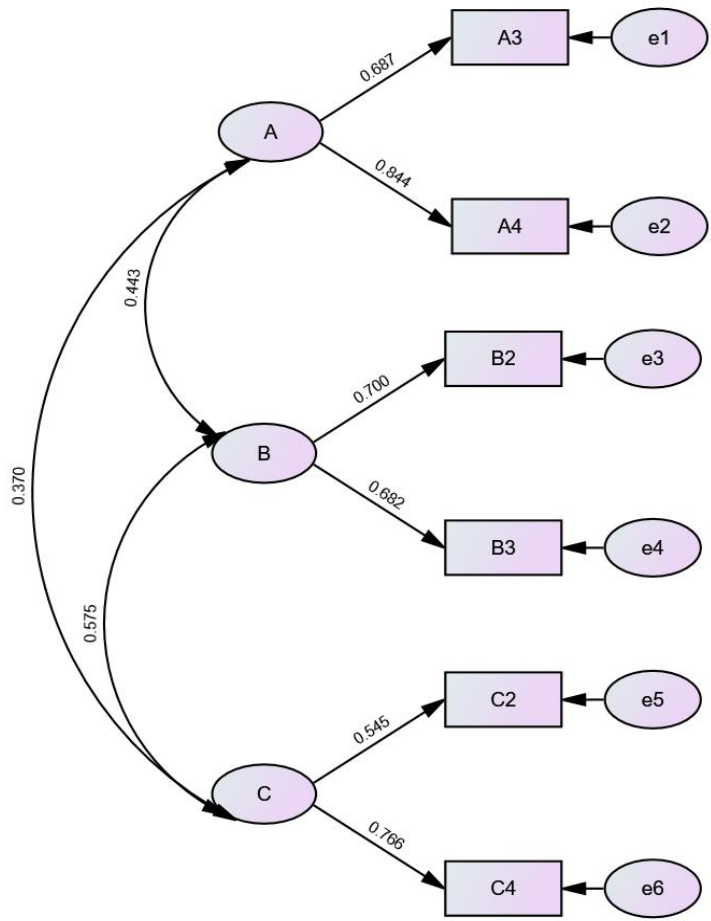

Figure 3 The Medication Literary Scale–Short Form: Confirmatory factor analysis : Three-factor model :A: functional ML, B:communicative ML, and C:critical ML

## 3.6 Reliability and Validity of the Simplified Medication Literacy Scale (MLS-SF)

The reliability and validity of the simplified Medication Literacy Scale (MLS-SF) was tested using sample 2. The analysis showed that the Cronbach's α coefficient of MLS-SF was 0.765. The item loadings ranged from 0.55 to 0.90, and the CFA model fit indices were as follows ($\chi^2$/df=5.11, RMSEA=0.056, GFI=0.990, AGFI=0.966,

NFI=0.984, IFI=0.987, CFI=0.987), suitable for factor analysis[36].In sample2, Bartlett's test of sphericity and KMO measure were conducted for MLS-SF. The Bartlett's spherical test value for MLS-SF was 1902.30 ($P<0.001$), and the KMO measure was 0.65, indicating that factor analysis could be performed. Principal component analysis was used without specifying the number of extracted factors, retaining factors with an eigenvalue greater than 1, and factor rotation method employed the Promax oblique rotation. The results showed that a total of 3 factors were retained, accounting for 66% of the cumulative variance explained. This validated its multidimensionality from the perspective of CTT. After oblique factor rotation, all 6 item factor loadings were greater than 0.65, indicating a high correlation between each item and its corresponding factor. Additionally, the results of the study showed that the I-CVI of the MLS-SF ranged from 0.875 to 1.000, and the S-CVI was 0.952, indicating that the Medication Literacy Short Form scale(MLS-SF) has good content validity.(Online Supplement)

The IRT analysis results showed reasonable discrimination and ICC distribution, with discrimination parameters ranging from 2.31 to 3.41, and difficulty parameters from -2.98 to 2.07. The average information curve is shown in the ICC figure (appendix), and there were no items with information less than 1 across the $\theta$ levels of -2 to 2, indicating a good distribution of item information.

## Discussion

This study developed short version of Medication Literacy Scale (MLS-14) and investigated the psychometric properties of the simplified scale MLS-SF in Chinese undergraduate population via both classical test theory and item response theory approaches. We conducted a comprehensive assessment of the reliability and validity of the simplified version of the scale. The MLS-SF developed in this study has a small number of questions, reducing the burden of filling out the questionnaire on the respondents and improving the questionnaire's usefulness and convenience.

Classical test theory has been commonly used in previous simplified studies. On the one hand, classical Test Theory (CTT) is a widely recognized and utilized psychometric approach with a well-established history[19].In other hand ,it interprets test scores as a combination of true and error scores, offering a comprehensive system for calculating reliability, validity, and other key parameters, thus enhancing the objectivity of testing. CTT and estimation methods are user-friendly, and its standardization techniques help minimize measurement errors. However, CTT estimates are sample-dependent and may not account for individual variability in error, a limitation addressed by Item Response Theory (IRT)[39]. IRT provides sample-invariant parameter estimates, allowing for precise measurement tailored to individual levels and accurately estimating item and test errors. In IRT, the estimation of ability parameters and item parameters is independent of the sample, and it can accurately estimate the measurement error of each item and test for different levels of individuals [40]. Therefore, more and more scholars suggest using both CTT and IRT to analyze the psychometric characteristics of measurement tools to comprehensively evaluate the applicability of measurement tools [41].

In this study, both CTT and IRT methods were used to obtain more accurate data. Before simplification, this study used confirmatory factor analysis to validate the structural validity of the original scale in sample 1. The results indicated that the Medication Literacy Scale (MLS-14) has a three-factor structure with good structural validity. In the process of scale simplification based on Classical test theory, this study also streamlined the Medication Literacy Scale according to the methods of four classical test theories commonly used in item analysis, and formed a 6-item short version of the scale (ML-SF) after deleting it. In the IRT, this study established a multi-dimensional GRM model based on the conclusions of the CTT three-factor model, and analyzed the discrimination parameters, local parameters, and DIF for each item. Combining CTT and IRT formed the six-item MLS-SF.
This study employed both CTT and IRT methods to enhance the accuracy of data. Confirmatory factor analysis was used to validate the three-factor structure of the original Medication Literacy Scale (MLS-14), demonstrating good structural validity.

The scale was then streamlined using CTT methods to create a six-item short form (MLS-SF). IRT was subsequently applied to develop a multidimensional model, analyzing item discrimination and difficulty, and ensuring the final MLS-SF combined the strengths of both theories for a comprehensive psychometric evaluation. The results of reliability analysis have shown that the high internal consistency reliability, as evidenced by a Cronbach's α coefficient of 0.765, suggest that the MLS-SF items are homogenous and measure the same construct of medication literacy. These reliability coefficients are within the acceptable range for new scales. The confirmatory factor analysis (CFA) results further support the construct validity of the MLS-SF, with all fit indices meeting the ideal criteria, indicating that the scale's structure is consistent with the hypothesized three-factor model. It shows that the MLS-SF can effectively measure different dimensions of medication literacy. The results of this study showed that the I-CVI ranged from 0.875 to 1.000 and the S-CVI was 0.952.It showed that the content validity of the MLS-SF was good and the entries were effective in measuring the level of medication literacy level.

The IRT analysis provided additional insights into the quality of the items. The discrimination parameters of MLS-SF were found to be within an acceptable range (2.31-3.41), indicating that the items can effectively differentiate between individuals with varying levels of medication literacy. The difficulty parameters also fell within a reasonable range (-2.98 to 2.07), suggesting that the items are neither too easy nor too challenging for the target population. The item information curves (ICCs) showed good distribution across the ability levels, with no items having information less than 1, which is indicative of the items' ability to provide information across the spectrum of medication literacy.It shows that the scale can provide differentiation and information for individuals with different levels of medication literacy.

The conciseness and psychometric properties of the MLS-SF make it a powerful tool for assessing and enhancing medication safety and health literacy among college students. More representative, the short version of the scale retained the most representative questions from a large number of scales and better reflected the theoretical structure of the Medication Literacy Scale, so as to target the evaluation of the main aspects of Medication literacy measurements, and reduces redundancy of the questions. In future medication education and public health interventions, this scale can serve as an effective assessment tool to identify individuals with lower levels of medication literacy and provide them with tailored education and support. Additionally, the brevity of the scale also makes it suitable for use in large-scale epidemiological studies to quickly collect data and evaluate the general level of medication literacy.

Two samples were included in this study to mutually validate the findings. The streamlined Medication Literacy Scale has fewer items, shorter response time, and lower response difficulty than commonly used assessment questionnaires in China, making it more suitable for use with all-age populations or in comprehensive questionnaires.

## Limitation

This study adopts the classical test theory and item response theory as the basis, and strictly adheres to the theoretical principles of scale simplification, but there are still some limitations. Firstly, our research sample was limited to college students in China, which may not fully represent other age groups or individuals from different cultural backgrounds. Therefore, future research needs to validate the applicability and measurement equivalence of the MLS-SF in a broader population. Secondly, the study employed a cross-sectional research design; future studies should conduct longitudinal research to assess the stability and long-term validity of the scale. Additionally, there may be a social desirability bias, where respondents might adjust their answers to fit societal expectations. Lastly, as a new measurement tool, the MLS-SF still requires more empirical research to accumulate evidence for its reliability and validity.

Future research should consider the following aspects: firstly, assess the applicability of the MLS-SF in different populations (such as the elderly, children, and individuals from different countries and regions); secondly, conduct longitudinal studies to examine the stability and long-term validity of the scale; thirdly, explore the application effects of the scale in different educational and intervention measures, and how to improve medication literacy levels through interventions; finally, consider developing and validating an electronic version of the scale for use in digital environments. Future studies should also consider retest intervals in order to obtain stronger evidence of the reliability of the scale.

## Conclusion

MLS-SF is a effective tool with a development perspective for the assessment of medication literacy among college students. Its brevity, psychometric soundness, and practicality position it well for integration into research and practice aimed at improving medication safety and health literacy. As the field of medication literacy continues to evolve, the MLS-SF may serve as a valuable instrument for evaluating and enhancing the knowledge and skills necessary for safe and effective medication use.